\documentclass[twocolumn,secnumarabic,amssymb, nobibnotes, aps, pre]{revtex4-2}

\usepackage{framed}
\usepackage{amsfonts}
\usepackage{amsmath}
\usepackage{graphicx}
\usepackage{color}

\usepackage[hidelinks]{hyperref}
\usepackage{xcolor}
\hypersetup{
	colorlinks,
	linkcolor={blue!80},
	citecolor={blue},
	urlcolor={blue!80!black}
}

\begin{document}
	\title{Dissipative solitons characterization in singly resonant optical parametric oscillators: a variational formalism }
	
	\author{Pedro Parra-Rivas}
	\email{pedro.parra-rivas@uniroma1.it}
	\affiliation{
		Dipartimento  di  Ingegneria  dell’Informazione, Elettronica  e  Telecomunicazioni,
		Sapienza  Universit{\'a}  di  Roma, via  Eudossiana  18, 00184  Rome, Italy\\
	}


	\begin{abstract}
 In this work, the emergence of single-peak temporal dissipative solitons in singly-resonant degenerate optical parametric oscillators is investigated analytically. Applying the Kantarovich optimization method, through a Lagrangian variational formalism, an approximate analytical soliton solution is computed using a parameter-dependent ansatz. This permits to obtain analytical estimations for the dissipative soliton energy, peak power, and existence boundaries, which are of great value for experimentalist. To confirm the validity of this procedure, these analytical results are compared with a numerical study performed in the context of pure quadratic systems, showing a good agreement.  
	\end{abstract}

\maketitle

\section{Introduction}
Solitons are localized nonlinear wave packets, with particle-like features, which propagate without suffering any shape modification
\cite{dauxois_physics_2006}. Although first reported in 1834 \cite{dauxois_physics_2006}, the term soliton was coined much later in 1965 by Zabuski and Kruskal while studying pulse interactions in collisionless plasmas \cite{PhysRevLett.15.240}.
In conservative systems, solitons arise due to the balance between spatial coupling and nonlinearity, and have been reported in a bast variety of natural context ranging from hydrodynamics to plasma physics \cite{dauxois_physics_2006,kivshar_optical_2003,10.1063/9780735425118}. In the context of optics, for example, temporal solitons may form in Kerr nonlinear single-pass wave-guides (e.g., optical fibers) where the spatial coupling role is played by dispersion, while Kerr nonlinearity is responsible for the self-phase modulation \cite{kivshar_optical_2003}.

Solitons may also arise in systems which are far from the thermodynamic equilibrium, where there exist a continue exchange of energy with the surrounding media. In this context, the so called {\it dissiaptive solitons} (DSs) \cite{akhmediev_dissipative_2005} can form if together with the balance between nonlinearity and spatial coupling, the system energy dissipation is compensated by external or internal driving. Here, DSs appears as isolated states ({\it attractors}), in contrast to conservative ones which form continues state families \cite{akhmediev_dissipative_2005}.

DSs arise naturally in nonlinear optical cavities where confined light can propagate indefinitely. In this context, DSs were first demonstrated experimentally in laser cavities \cite{taranenko_spatial_1997,weiss_solitons_1999}, and later in semiconductor externally driven microcavities \cite{barland_cavity_2002}. In all these cases, DSs are two-dimensional objects forming in the transverse plane to the light propagation direction.  

One-dimensional (temporal) DSs were shown experimentally in fiber cavities and microresonators, where localization takes place along the propagation direction \cite{leo_temporal_2010,herr_temporal_2014}. These states have been proposed for different technological applications including all-optical buffering \cite{leo_temporal_2010} and frequency comb generation \cite{herr_temporal_2014,brasch_photonic_2016}.

DSs have been extensively studied in diffractive cavities in one- and two-dimensions \cite{staliunas_localized_1997,longhi_localized_1997,trillo_stable_1997,staliunas_spatial-localized_1998,oppo_domain_1999,oppo_characterization_2001,rabbiosi_new_2003,etrich_solitary_1997,Trillo:98,PhysRevE.60.R3508}, and more recently, their temporal counterparts have been also predicted and analyzed in pure quadratic dispersive cavities, either in the context of cavity enhanced second-harmonic generation \cite{hansson_quadratic_2018,villois_frequency_2019,arabi_localized_2020,PhysRevA.104.063502,lu_two-colour_2023}, or in degenerate optical parametric oscillators (DOPO), in singly \cite{PhysRevApplied.13.044046,PhysRevResearch.4.013044} and doubly resonant configurations \cite{parra-rivas_frequency_2019,parra-rivas_localized_2019,parra-rivas_parametric_2020,nie_quadratic_2020-2}. Moreover, the implications of Kerr competing nonlinearities have also been analyzed \cite{villois_soliton_2019,MasArabi:23}, and parametric Kerr DSs recently demonstrated experimentally \cite{englebert_parametrically_2021}.    

In this work, I study the formation of (single-peak) temporal DSs in singly-resonant DOPO, hereafter SR-DOPO, composed by one section with quadratic nonlinearity and another one formed by a Kerr material \cite{englebert_parametrically_2021,MasArabi:23}. The approach that I will follow relies on the Lagrangian variational formulation and on the Kantarovich optimization method for dissipative systems \cite{chavez_cerda_variational_1998}. This machinery allows, providing a suitable parameter-dependent solution ansatz, the computation of DSs analytical approximations which agree quite well with the exact numerical solutions studied in Ref.~\cite{PhysRevResearch.4.013044}. 


The paper is organized as follow. In Section~\ref{sec:1} the mean-field model describing SR-DOPO cavities is introduced. After that, Section~\ref{sec:2} is devoted to the presentation of the variational formulation associated with this model. Section~\ref{sec:3} applies the Kantarovich optimization method for computing an approximation DS solution. Later, in Section~\ref{sec:4} these analytical results are compared with numerical ones obtained by means of path-continuation algorithms. Finally, Section~\ref{sec:5} draws the final conclusions of this work.   
\section{The mean-field model}\label{sec:1}
\begin{figure}[!t]
	\centering
	\includegraphics[scale=1]{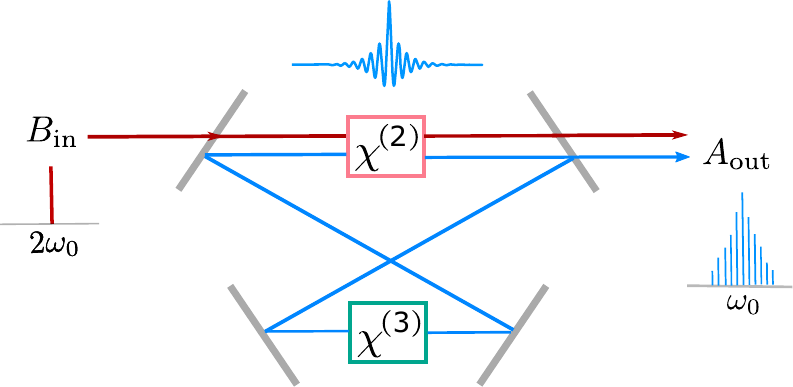}
	\caption{Schematic representation of a SR-DOPO with a Kerr nonlinear section. The cavity contains a nonlinear quadratic $(\chi^{(2)})$ medium and a section with a Kerr nonlinear one $(\chi^{(3)}).$ The cavity is driven by a continuous-wave field $B_{in}$ at frequency $2\omega_0$. Here only the generated signal field $A$ at frequency $\omega_0$ resonates, while the pump field $B$ leaves the cavity at each round-trip.  }
	\label{fig1}
\end{figure}

The system that we consider in this analysis is a SR-DOPO composed by two nonlinear sections: a quadratic material ($\chi^{(2)}$) and a Kerr one ($\chi^{(3)}$), as depicted in Fig.~\ref{fig1}. 
The pump field $B_{in}$ at frequency $2\omega_0$  is injected in the quadratic material where, through phase-matched
parametric downconversion, the signal at frequencies around $\omega_0$ 
is amplified. After propagating in the $\chi^{(2)}$-material, the pump is extracted from
the cavity, so that only the signal resonates, interacting with the Kerr medium.

\begin{figure}[!t]
	\centering
	\includegraphics[scale=0.95]{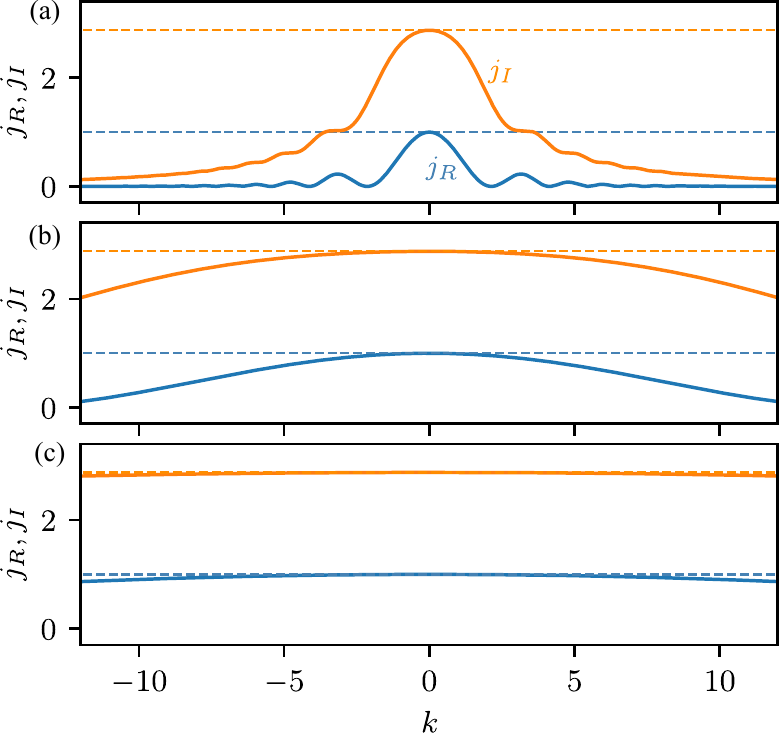}
	\caption{Modification of the kernel Fourier transform shape with $\eta_2$ for $(d,\varrho)=(0,-4)$. In (a) $\eta_2=0.5$, in (b) $\eta_2=0.01$, and in (c) $\eta_2=0.001$.The horizontal lines correspond to the values $\kappa=j_R(0)=1$ (blue) and $\nu_2=j_I(0)$ (orange).  }
	\label{fig2}
\end{figure}

In the mean-field approximation, the Ikeda map describing this cavity can be reduced to the following dimensionless partial differential equation with nonlocal nonlinearity \cite{mosca_modulation_2018,englebert_parametrically_2021,MasArabi:23}
\begin{equation}\label{Eq.1st}
	u_t=-(\alpha+i\delta)u-i\eta_1 u_{xx}+i\nu_1|u|^2u+[\mu -\rho(u^2\otimes J)]u^*,
\end{equation}
where  $t$ is the slow time, $x$ the fast time, $u$ is the slowing varying envelope of the signal electric field circulating in the cavity and $u^*$ its complex conjugate, $\delta$ is the normalized phase detuning from the closest cavity resonance, $\eta_1$ represents the normalized group velocity dispersion of $u$ at $\omega_0$, and $\mu$ is the normalized amplitude of the driving field. The term $u^2\otimes J$ represents the nonlocal nonlinearity defined through the convolution ($\otimes$) between $u^2$ and the nonlocal kernel
\begin{equation}
	J(x)=\frac{1}{2\pi}\int_{-\infty}^{\infty}j(k)e^{-ik x}dx.
\end{equation}
The Fourier transform of $J$, namely $\mathcal{F}[J](k)\equiv j(k)=j_R(k)+ij_I(k)$ is defined through the expressions

\begin{equation}
	j_R(k)=\frac{{\rm sinc}^2(Z(k)/2)}{2c(\varrho)},\qquad
	j_I(k)=\frac{{\rm sinc}(Z(k))-1}{c(\varrho)Z(k)},
\end{equation}
where
\begin{equation}
	Z(k)\equiv\varrho-dk-\eta_2 k^2,\qquad c(\varrho)\equiv\frac{1}{2}{\rm sinc}^2(\varrho/2),
\end{equation}
and $\varrho$, $d$ and $\eta_2$ are the normalized phase mismatch, group velocity mismatch or walk-off, and group velocity dispersion of the pump field, respectively \cite{PhysRevResearch.4.013044}. This nonlocal nonlinearity acts as pump ($\mu$) depletion, and may introduce important modification on the DS dynamics and stability \cite{englebert_parametrically_2021,MasArabi:23}.  Here, $j_R(k)$ mirror the dispersive two-photon absorption, while  the contribution related with $j_I(k)$ produces a phase-shift, similar to the Kerr effect \cite{PhysRevApplied.13.044046,PhysRevResearch.4.013044}. In what follows, we study the formation of DSs in the absence of walk-off ($d=0$). Figure~\ref{fig2} shows the modification of $j_R(k)$ and $j_I(k)$ for three decreasing values of $\eta_2$ for $\varrho=-4$ \cite{PhysRevResearch.4.013044}.

The previous mean-field model can describe two type of systems. 
When $\nu_1=0$ and $\rho=1$, Eq.~(\ref{Eq.1st}) models pure quadratic SR-DOPOs, where the nonlinearity is only of second order. This equation has been used to describe the formation of quadratic frequency combs \cite{mosca_modulation_2018} and DS formation in different regimes of operation \cite{PhysRevApplied.13.044046,PhysRevResearch.4.013044}.
When $\nu_1\neq0$, Eq.~(\ref{Eq.1st}) can model a SR-DOPO with a Kerr nonlinear section like the one illustrated in Fig.~\ref{fig1}. In that case, $\rho$ represents the ratio between the quadratic and cubic nonlinearities. This last configuration has been recently considered to study the effect of pump depletion in Kerr soliton formation \cite{MasArabi:23}.  Finally, when $\rho=0$ this model describes a SR-DOPO with just a Kerr section, when the pump depletion is neglected \cite{Longhi:95,englebert_parametrically_2021}. 


\section{The variational formulation}\label{sec:2}
Equation~(\ref{Eq.1st}) can be derived following a variational approach \cite{wiggins_introduction_2003}. This procedure requires the definition of a Lagrangian density $\mathcal{L}$ describing the conservative dynamics of the field $u$:
\begin{equation}
	\mathcal{L}\equiv \frac{i}{2}(u u^*_t-u^*u_t)-\eta_1 u_xu^*_x-\frac{1}{2}\nu_1u^2u^{*2}+\delta u u^*, 
\end{equation}
and a Rayleigh dissipative functional which describes the non-conservative effects \cite{chavez_cerda_variational_1998,PhysRevE.100.022201,4610001,Yi:16,ankiewicz_dissipative_2007}. In our case,  the Rayleigh functional is defined as
\begin{equation}
	\mathcal{R}=\mathcal{R}_\alpha+\mathcal{R}_\mu+\mathcal{R}_\rho,
\end{equation}
with the components
\begin{equation}
	\mathcal{R}_\alpha\equiv i\alpha(u u^*_t-u^*u_t),
\end{equation}
\begin{equation}
	\mathcal{R}_\mu\equiv i\mu(uu_t-u^*u_t^*),
\end{equation}
\begin{equation}
	\mathcal{R}_\rho\equiv i\rho\left[(u^2\otimes J)u^*u_t^*-(u^2\otimes J)^*uu_t\right],
\end{equation}
associated with losses, external driving and long-range non-local term, respectively. 
With these definitions, Eq.~(\ref{Eq.1st}) can be recovered from the equation
\begin{equation}\label{GEL}
	i\left(\mathcal{E}+\mathcal{Q}\right)=0,
\end{equation}
where $\mathcal{E}$ is the Euler-Lagrange term
\begin{equation}
	\mathcal{E}\equiv\frac{d}{dt}\left(\frac{\partial \mathcal{L}}{\partial u^*_t}\right)+\frac{d}{dx}\left(\frac{\partial \mathcal{L}}{\partial u^*_x}\right)-\frac{\partial\mathcal{L}}{\partial u^*},
\end{equation}
and 
\begin{equation}
	\mathcal{Q}\equiv\mathcal{Q}_\alpha+\mathcal{Q}_\mu+\mathcal{Q}_\rho=\frac{\partial \mathcal{R}_\alpha}{\partial u^*_t}+\frac{\partial \mathcal{R}_\mu}{\partial u^*_t}+\frac{\partial \mathcal{R}_\rho}{\partial u^*_t}
\end{equation}
represents all the non-conservative terms.
This is the general approach when dealing with dissipative systems \cite{chavez_cerda_variational_1998,PhysRevE.100.022201,PhysRevA.96.013838}, although there are other equivalent alternatives \cite{PhysRevLett.83.2568}.

\subsection*{The parametrically forced Ginzburg-Landau approximation}
Although all the variational approach previously discussed is completely general, the complexity of the nonlocal kernel $J$, makes intractable the computation of an analytical variational approximation of the DS solution. 
To avoid this inconvenience, in what follows, I am going to work in the broadband limit where the approximation
$\mathcal{F}[J](k)\approx\mathcal{F}[J](0)$ can be considered \cite{PhysRevApplied.13.044046}. This limit can be reached, for example, when decreasing $\eta_2$, as shown in Fig.~\ref{fig2}: for $\eta_2=0.5$ [see Fig.~\ref{fig2}(a)], $j_R$ and $j_I$ are quite localized in $k$
(i.e., narrowband); by decreasing $\eta_2$ [see Fig.~\ref{fig2}(b) for $\eta_2=0.01$] $j_R$ and $j_I$ broaden; and for very small $\eta_2$ they are almost constant [see Fig.~\ref{fig2}(c) for $\eta_2=0.001$]. 

As described in \ref{app2}, this approximation leads to, 
\begin{equation}
	\mathcal{R}_\rho\approx -2\rho|u|^2{\rm Im}\left[(\kappa+i\nu_2)uu^*_t\right], 
\end{equation}
with 
\begin{equation}
	\kappa\equiv j_R(0)=1 ,\qquad \nu_2\equiv j_I(0)=\frac{{\rm sinc}(\varrho)-1}{c(\varrho)\varrho}.
\end{equation}


After this consideration, Eq.~(\ref{Eq.1st}) leads to the local parametrically forced Ginzburg-Landau equation (PFGLE) 
$$	u_t=-(\alpha+i\delta)u-i\eta_1 u_{xx}+\mu u^*+\left[\rho\kappa+i(\nu_1+\rho\nu_2)\right]|u|^2u.$$
The case with $\rho=0$, where nonlinearity reduces to a pure Kerr term , was originally proposed by Longhi \cite{Longhi:95}, and utilized recently for modeling the formation of parametric Kerr DSs \cite{englebert_parametrically_2021}. It is worth to mention that when temporal walk-off is very large $d\gg1$, other kernel approximations allow the analytical integration of the convolution term as described in Ref.~\cite{MasArabi:23}.
\section{The Kantarovich optimization method: the dissipative soliton solution}\label{sec:3}
This section is devoted to the derivation of  a reduced effective dynamical system able to describe the main DSs features. To do so, I apply the Kantarovich method, a generalization of the Rayleigh-Ritz optimization method \cite{chavez_cerda_variational_1998,ankiewicz_dissipative_2007,kaup_variational_1995}. Given a parameter-dependent DS solution ansatz, this method follows a variational approach to optimize such parameters to fit the DS exact solution. 
Therefore, the most important part in this procedure is the election of a proper solution ansatz. 

Here, based on previous works \cite{Longhi:95,PhysRevE.53.5520}, I propose a two-parameter DS ansatz of the form
\begin{equation}\label{ansatz}	u(x,t)=q_1(t)B[x,q_1(t)]e^{iq_2(t)},
\end{equation}
with 
$	B[x,q_1(t)]\equiv {\rm sech}[aq_1(t)x]$,
where $q_1$ and $q_2$ correspond to the time-dependent amplitude and phase, respectively. The Kantarovich generalization of the Rayleigh-Ritz method consist basically in allowing these parameters depend on time. Here, the factor $a$ relates the amplitude of the soliton with its width, and to determine it, I have considered the conservative soliton formation condition as described in \cite{agrawal_applications_2008}. This condition leads to 
$$a=\sqrt{\frac{\nu_1+\rho\nu_2}{2}}.$$


The temporal evolution of a DS of this form can be described using the reduced effective dynamical system derived from the Lagrangian and Rayleigh functions
\begin{equation}\label{LagInt}
	L(q_1,\dot{q}_1,q_2,\dot{q}_2)\equiv\int_{-\infty}^{\infty}\mathcal{L}(u,u_{xx},u_t)dx,
\end{equation}
\begin{equation}\label{RayInt}
	R(q_1,\dot{q}_1,q_2,\dot{q}_2)\equiv\int_{-\infty}^{\infty}\mathcal{R}(u,u_t)dx.
\end{equation}

After performing the integration, the effective Lagrangian becomes
\begin{equation}\label{effecLag}
	L=\frac{2}{a}\left[q_1\dot{q}_2+\delta q_1-\frac{1}{3}\left(a^2\eta_1+\nu_1\right)q_1^3\right],
\end{equation}
and the Rayleigh function reads
$R=R_\alpha+R_\mu+R_\rho,$
with 
\begin{equation}\label{effecRay1}
	R_\alpha=\frac{4\alpha}{a}q_1 \dot{q}_2, 
\end{equation}
\begin{equation}\label{effecRay2}
	R_\mu=-\frac{2\mu}{a}\left[{\rm sin}(2q_2)\dot{q}_1+2q_1{\rm cos}(2q_2)\dot{q}_2\right],
\end{equation}
and 
\begin{equation}\label{effecRay3}
	R_\rho=\frac{2\rho}{a}\left(\frac{4}{3}\kappa  q_1^3\dot{q}_2-\nu_2q_1^2\dot{q}_1\right).
\end{equation}

From these functions, a 2D dynamical system, describing the evolution of $(q_1,q_2)$, can be derived from the dissipative version of the Euler-Lagrange equations:
$$	\frac{d}{dt}\left(\frac{\partial L}{\partial \dot{q}_j}\right)-\frac{\partial L}{\partial q_j}+Q^{(j)}_{\alpha}+Q^{(j)}_{\mu}+Q^{(j)}_{\rho}=0,\qquad Q^{(j)}_{\xi}\equiv\frac{\partial R_{\xi}}{\partial \dot{q}_j},$$
for $j=1,2$.
These equations yield the system (see derivation in \ref{app4}):
\begin{equation}\label{red_eq1}
	\dot{q}_1=-2\alpha q_1-\frac{4}{3}\rho\kappa q_1^3+2\mu q_1{\rm cos}(2q_2),
\end{equation}
\begin{equation}\label{red_eq2}
	\dot{q}_2=-\delta+(a^2\eta_1+\nu_1-\rho\nu_2)q_1^2-\mu{\rm sin}(2q_2).
\end{equation}
\begin{figure*}[!t]
	\centering
	\includegraphics[scale=1.05]{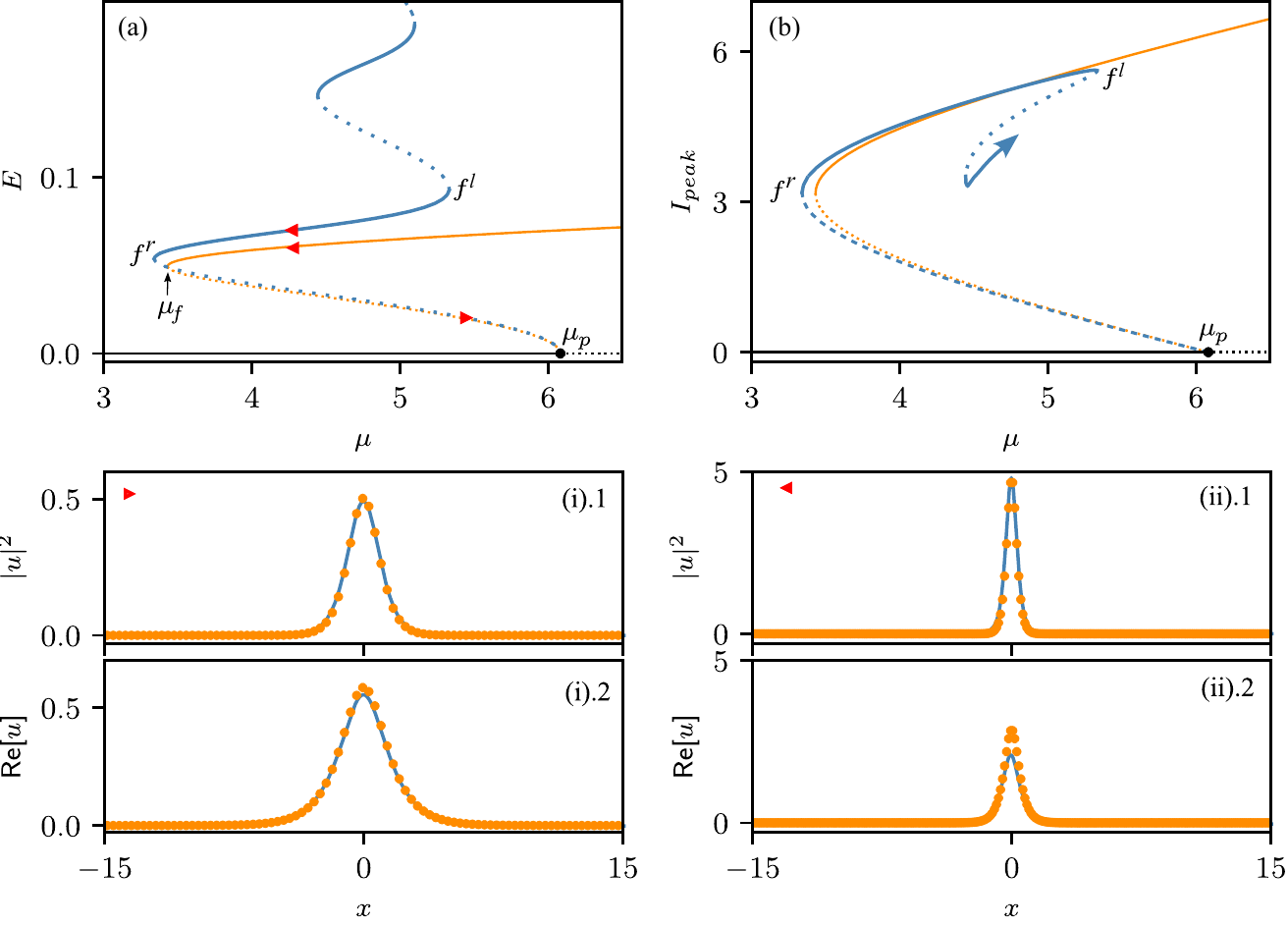}
	\caption{Comparison between numerical (blue) and analytical (orange) bifurcation diagrams and DS solutions. Panel (a) shows the DS energy as a function of $\mu$ for $\delta=-6$. Panel (b) shows the peak intensity $I_{peak}$ of the DSs shown in (a) as a function of $\mu$. Solid (dashed) lines represent stable (unstable) states. Examples of an unstable (see {\color{red}$\blacktriangleright$}) and stable DS state (see {\color{red}$\blacktriangleleft$}) are shown in the panels below. Here,$(\eta_2,\varrho)=(0.01,-4)$. 
	}
	\label{fig3}
\end{figure*}
The fixed points of Eqs.~(\ref{red_eq1}) and (\ref{red_eq1}) satisfy $(\dot{q}_1,\dot{q}_2)=(0,0)$, and their amplitude component
leads to the solutions 
\begin{equation}\label{sol_q1}
	q_1=0, \qquad	q_1^{(\pm)}=\pm\sqrt{\frac{\delta C_2-\alpha C_1\pm \sqrt{\Lambda}}{C_1^2+C_2^2}},
\end{equation}
with 
\begin{equation}
	\Lambda\equiv (\alpha C_1-\delta C_2)^2-(C_1^2+C_2^2)(\alpha^2+\delta^2-\mu^2),
\end{equation}
$C_1\equiv2\rho\kappa/3$, and $C_2\equiv a^2\eta_1+\nu_1-\rho\nu_2$, as described in \ref{app5}.
Therefore, the system has the trivial solution $q_1=0$, which corresponds to the continuous wave state \cite{PhysRevResearch.4.013044}, and the non-trivial solutions $q_1^{(+)}$, and  $q_1^{(-)}$, which correspond to two coexisting DS states.

When these two solution meet [i.e., when $q_1^{(+)}=q_1^{(-)}$], $\Lambda=0$, and 
a fold bifurcation occurs at 
\begin{equation}
	q^f_1=\pm\sqrt{\frac{\delta C_2-\alpha C_1}{C_1^2+C_2^2}},	
\end{equation}
\begin{equation}\label{fold}
	\mu_f^2=\alpha^2+\delta^2-\frac{(\alpha C_1-\delta C_2)^2}{C_1^2+C_2^2}.	
\end{equation}
In contrast, if $q_1^{(-)}$ collides with $q_1=0$,
a pitchfork bifurcation takes place at
\begin{equation}\label{pitch}
	\mu_p^2=\alpha^2+\delta^2.
\end{equation}
Note that the same last expression can be obtained by means of multi-scale perturbation techniques as shown in \cite{PhysRevResearch.4.013044}.

The phase contribution to the DS solution leads to (see \ref{app5}):
\begin{equation}
	{\rm sin}(q_2)=\sqrt{\frac{\mu-\alpha-2\rho\kappa q_1^2/3}{2\mu}}
\end{equation}
\begin{equation}
	{\rm cos}(q_2)=\sqrt{\frac{\mu+\alpha+2\rho\kappa q_1^2/3}{2\mu}},
\end{equation}
which depends on $q_1$.

With all these expressions, the intensity profiles corresponding the previous DS solutions read 
\begin{equation}
	I^{(\pm)}(x)\equiv|u^{(\pm)}(x)|^2=[q_1^{(\pm)}]^2{\rm sech}^2\left(\sqrt{\frac{\nu_1+\rho\nu_2}{2}}q_1^{(\pm)}x\right),
\end{equation}
while their real and imaginary parts are
\begin{equation}
	{\rm Re}[u^{(\pm)}(x)]=|u(x)^{(\pm)}|\sqrt{\frac{\mu+\alpha+2\rho\kappa q_1^2/3}{2\mu}},
\end{equation}
\begin{equation}
	{\rm Im}[u^{(\pm)}(x)]=|u(x)^{(\pm)}|\sqrt{\frac{\mu-\alpha-2\rho\kappa q_1^2/3}{2\mu}}.
\end{equation}
These expressions allow the computation of important DS magnitudes such as the peak intensity 
\begin{equation}\label{peak}
	I_{peak}^{(\pm)}\equiv I^{(\pm)}(0),
\end{equation}
and the DS energy
\begin{equation}\label{Ener}
	E\equiv\int_{-\infty}^\infty|u^{(\pm)}(x)|^2dx=\frac{2\sqrt{2}q_1^{(\pm)}}{\sqrt{\nu_1+\rho\nu_2}}.
\end{equation}

\section{A particular case of study: the pure quadratic SR-DOPO}\label{sec:4}
To confirm the previous analytical results, I compare them with those obtained numerically in previous studies on the pure quadratic SR-DOPO scenario \cite{PhysRevResearch.4.013044}. Thus, in what follows, I fix $\nu_1=0$, $\rho=1$, and $\alpha=1$. This election leads to the coefficients 
\begin{equation}
	a=\sqrt{\nu_2/2}, \qquad C_1=2/3, \qquad C_2=(\eta_1-2)\nu_2/2.
\end{equation}
Moreover, I also choose $\eta_1=1$, $\varrho=-4$, and $\eta_2=0.01$.
For this regime of parameters, DSs undergo a bifurcation structure known as {\it collapsed snaking} \cite{knobloch_homoclinic_2005} which has been characterized in detail in Ref.~\cite{PhysRevResearch.4.013044}. A portion of this curve, computed by means of path-numerical continuation methods \cite{allgower_numerical_1990}, is depicted in Fig.~\ref{fig3}(a), where the DSs 
energy is plotted as a function of $\mu$ for $\delta=-6$ [see blue curve]. The stability of these states has been computed numerically and is depicted using solid (dashed) lines for stable (unstable) states. 

The DS states on the first stable branch, limited by the fold points $f^{l,r}$, correspond single-peak DSs with a {\it sech}-shape. The intensity and real profiles of two single-peak DSs on the stable and unstable solution branches are illustrated  
in Fig.~\ref{fig3}(i), (ii) for $\mu=5.24$ and $\mu=4.27$, respectively (see blue curves). The modification of the DS peak intensity $I_{peak}$ along this diagram is also plotted in Fig.~\ref{fig3}(b).

At this stage, there is enough information to compare these numerical results with the analytical ones obtained in Section~\ref{sec:3}.
The analytical prediction of the stable and unstable DS solutions are plotted using orange dots in Figs.~\ref{fig3}(i) and \ref{fig3}(ii). When regarding the intensity profiles [see Figs.~\ref{fig3}(i).1 and \ref{fig3}(ii).1], the agreement is excellent. By taking a look to the real profiles, however, one can see that while the agreement is considerably good for the unstable state [see Figs.~\ref{fig3}(i).2], it worsens for the stable one [see Figs.~\ref{fig3}(ii).2]. 

The good agreement shown by the analytical prediction of the intensity seems to persist for other values of $\mu$ as depicted in Fig.~\ref{fig3}(b) where the peak intensity of the DSs solutions [see Eq.~(\ref{peak})] is plotted as a function of $\mu$ using an orange line. When approaching $f^l$, the agreement between numerical and analytical results aggravates. With increasing $\mu$, however, the agreement is quite good for both the stable and unstable branches. Regarding the stable DSs, the variational approach does not predict the existence of the right fold $f^r$, and the analytical stable solution branch extends towards infinity. In contrast, for the stable states, 
the variational approach even describes the pitchfork point $\mu_p$ from where DSs arise.

It is worth to mention that, applying multi-scale perturbation theory, another approximate DS solution, complementary to the one found here, can be computed around $\mu=\mu_p$ \cite{PhysRevResearch.4.013044}. This solution reads
\begin{equation}
	u(x)=\left(\frac{\delta}{1-\mu_p}+i\right)\sqrt{\frac{2(\mu_p-\mu)}{c_3}}{{\rm sech}\left(\sqrt{\frac{\mu_p-\mu}{-c_1}}x\right)},	
\end{equation}
with 
$$c_1\equiv\frac{-2\eta_1\sigma}{1+\sigma^2}, \qquad c_3\equiv 2\sigma \nu_2(\varrho)-(\sigma^2-1),\qquad\sigma=\frac{\delta}{1-\mu_p}.$$

Figure~\ref{fig3}(a) shows the comparison between the numerical and analytical DS energy. The analytical prediction seems to agree quite well for most of the unstable solution branch. However, as approaching the $f^l$, the agreement worsens.


\begin{figure}[t]
	\centering
	\includegraphics[scale=1]{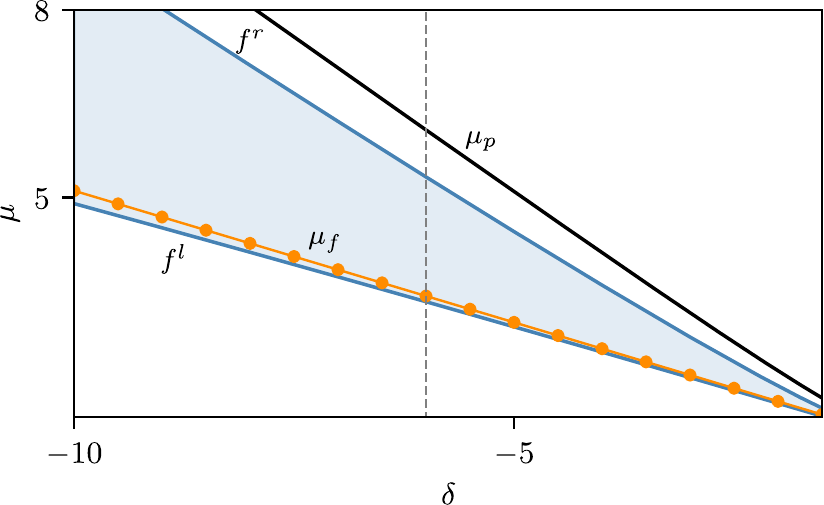}
	\caption{Phase diagram in the $(\delta,\mu)$-parameter space showing the main bifurcation lines of the system. The line with orange dots represents the analytical prediction for $f^l$ ($\mu_f$), and the black line corresponds to the pitchfork bifurcation line $\mu_p$. The blue shadowed area corresponds to the region of existence of DSs.  Here, $(\eta_2,\varrho)=(0.01,-4)$. }
	\label{fig4}
\end{figure}

Finally, the analytical prediction for the left fold and pitchfork points, respectively Eqs.~(\ref{fold}) and (\ref{pitch}), permits a comparison with their numerical counterparts. Figure~\ref{fig4} shows the $(\delta,\mu)$-parameter plane associated with the single-peak DS states. The blue area corresponds to the region of existence of such states, and has been determined numerically by a two-parameter continuation of the left and right fold points $f^{l,r}$ shown in Fig.~\ref{fig3}(a). The dashed vertical line in Fig.~\ref{fig4} corresponds to the diagrams plotted in Figs.~\ref{fig3}(a), (b).

The position of the analytically predicted left fold [see Eq.~(\ref{fold})] is plotted using orange dots. This prediction works quite well for low values of $|\delta|$. However, with increasing $|\delta|$, it separates from the numerical result. In contrast, the pitchfork bifurcation line obtained using the variational approach is exactly the same one obtained applying multi-scale perturbation theory \cite{PhysRevResearch.4.013044}. Therefore, the analytical predictions computed here can be used as good approximations for the DSs boundaries.

\section{Discussion and conclusions}\label{sec:5}
This work has presented an analytical study of single-peak DSs emerging in a SR-DOPO when considering two nonlinear sections: one $\chi^{(2)}$ and another $\chi^{(3)}$ . This system can be described by a generalized version of the parametrically forced Ginzburg-Landau equation which include a long-range interaction term [see Eq.~(\ref{Eq.1st}) in Section~\ref{sec:1}]. I show that equations of this type can be described using the variational formalism if non-conservative effects and the long-range interaction is taken into account through a Rayleigh dissipative functional \cite{chavez_cerda_variational_1998}. The use of the Rayleigh functional is a standard procedure when dealing with dissipative systems, as has been shown in different contexts \cite{chavez_cerda_variational_1998,4610001,Yi:16,tahOL19,kalashnikov_stabilization_2022,MasArabi:23}.

The main result of this work is the analytical computation of an approximate single-peak DS solution in terms of the Kantarovich optimization approach (see Section~\ref{sec:3}). Starting from a suitable parameter-dependent DS solution ansatz, this approach uses the variational formalism for reducing the mean-field model to a 2D dynamical system which describes the temporal evolution of the aforementioned parameters. Thus, the fixed points of this 2D system provide a good approximation for the original DS state. Furthermore, this approach yields the analytical estimation of some DS existence boundaries, and other features, such as the DS energy and peak intensity. 

The validity of these results is confirmed in Section~\ref{sec:4} by comparison with the numerical bifurcation analysis performed in the context of pure $\chi^{(2)}$ SR-DOPOs \cite{PhysRevResearch.4.013044}. The DS intensity profiles prediction and the their peak intensity match quite well the numerical results. In contrast, the agreement worsens when comparing the DSs energy. The analytical prediction of the DS boundary fold $f^l$ shows a quite good agreement, although it worsens a bit with increasing $|\delta|$. For the pitchfork bifurcation, however, this analysis leads to the same result than multi-scale perturbation methods \cite{PhysRevResearch.4.013044}.

The approach utilized here is very sensitive to the solution ansatz proposed. Therefore, the correct election of the ansatz is key for obtaining a good description of the solution under study. More complete ansätze, including for example temporal chirp, could lead to a better description of the system, although with a higher computational cost. In a future research, I will explore this line and seek for more accurate descriptions of the system.  


\section*{Acknowledgements}
I am thankful to Dr. Mas-Arabí for the fruitful conversations we had during the early stages of this work. Furthermore, I acknowledge support from the European Union’s Horizon 2020 research and innovation programme
under the Marie Sklodowska-Curie grant agreement no. 101023717.
\appendix


\section{Approximation of the Kernel}\label{app2}
In the broadband limit (see Fig.~\ref{fig2}), we can write  
$$	\mathcal{F}[J](k)\approx\mathcal{F}[J](0)=\kappa+i\nu_2,$$ 
where
\begin{eqnarray*}
	\kappa\equiv \mathcal{F}[J_R](0)=1,\qquad 	\nu_2\equiv \mathcal{F}[J_I](0)=\frac{{\rm sinc}(\varrho)-1}{c(\varrho)\varrho}.
\end{eqnarray*}
With this approximation, the convolution term reads
\begin{eqnarray*}
	u^2\otimes J\equiv \int_{-\infty}^{\infty}u^2(y)J(x-y)dy\approx \qquad\qquad\qquad\qquad\qquad\\\int_{-\infty}^{\infty}u^2(y)(\kappa+i\nu_2)\delta(x-y)dy=(\kappa+i\nu_2)u^2(x),
\end{eqnarray*}
and thus, the Rayleigh functional $\mathcal{R}_\rho$ becomes
\begin{eqnarray*}	\mathcal{R}_\rho=i\rho\left[(u^2\otimes J)u^*u_t^*-(u^2\otimes J)^*uu_t\right]=\qquad\qquad\qquad\\-2\rho {\rm Im}[(u^2\otimes J)u^*u_t^*]\approx -2\rho|u|^2{\rm Im}\left[(\kappa+i\nu_2)uu^*_t\right]. 
\end{eqnarray*}	
\section{The effective Lagrangian and Rayleigh function}\label{app3}
Inserting the ansatz defined by Eq.~(\ref{ansatz}), into the Lagrangian density one obtains
\begin{equation}
	\mathcal{L}= q_1^2\left(B^2\dot{q}_2-\eta_1 B_x^2-\frac{\nu_1}{2}q_1^2 B^4+\delta B^2\right),
\end{equation}
which once integrated [see Eq.~(\ref{LagInt})], yields the effective Lagrangian
\begin{equation}
	L=q_1^2\left(\dot{q}_2I_a+\delta I_a-\eta_1 I_b-\frac{\nu_1}{2}q_1^2I_c\right)
\end{equation}
with the integrals
$$	I_a\equiv\int_{-\infty}^\infty B^2dx=\frac{2}{aq_1},$$
$$	I_b\equiv\int_{-\infty}^\infty B_x^2dx=\frac{2aq_1}{3},$$
$$	I_c\equiv\int_{-\infty}^\infty B^4 dx=\frac{4}{3}\frac{1}{aq_1}.$$
Proceeding in a similar way with the different terms of the Rayleigh functional, the following expressions are obtained
$$\mathcal{R}_\alpha=-2\alpha{\rm Im}[uu^*_t]=2\alpha B^2 q_1^2 \dot{q}_2,$$ 
$$	\mathcal{R}_\mu=-2\mu\left(B^2q_1+q_1^2B\frac{\partial B}{\partial q_1}\right){\rm sin}(2q_2)\dot{q}_1\\-2\mu B^2q_1^2{\rm cos}(2q_2)\dot{q}_2,$$
$$	\mathcal{R}_\rho=2\rho\left[\kappa B^4 q_1^4\dot{q}_2-\nu_2q_1^3\left(B^4+q_1B^3\frac{\partial B}{\partial q_1}\right)\dot{q}_1\right].$$
After integration [see Eq.~(\ref{RayInt})], these terms leads to the Rayleigh functions
$$	R_\alpha=2\alpha I_a q_1^2 \dot{q}_2,$$
$$R_\mu=-2\mu\left[\left(I_aq_1+q_1^2I_d\right){\rm sin}(2q_2)\dot{q}_1+I_aq_1^2{\rm cos}(2q_2)\dot{q}_2\right],$$
$$	R_\rho=2\rho\left[\kappa I_c q_1^4\dot{q}_2-\nu_2q_1^3\left(I_c+q_1I_e\right)\dot{q}_1\right],$$
with the integrals 
$$	I_d\equiv\int_{-\infty}^\infty B\frac{\partial B}{\partial q_1}dx=-\frac{1}{a q_1^2},$$
$$I_e\equiv\int_{-\infty}^{\infty} B^3\frac{\partial B}{\partial q_1}dx=-\frac{1}{3aq_1^2}.$$

\section{The generalized Euler-Lagrange equations for the soliton parameters}\label{app4}
In this appendix the 2D dynamical system, composed by Eqs.~(\ref{red_eq1}) and (\ref{red_eq2}), is derived using the expressions for the effective Lagrangian and Rayleigh functions [see Eqs.~(\ref{effecLag}), (\ref{effecRay1})-(\ref{effecRay3})].

For $j=1$, 
$$	\frac{d}{dt}\left(\frac{\partial L}{\partial \dot{q}_1}\right)=0, \qquad \frac{\partial L}{\partial q_1}=\frac{2}{a}\left[\dot{q}_2+\delta-(a^2\eta_1+\nu_1)q_1^2\right],$$ 
$$	Q^{(1)}_\alpha=0, \qquad Q^{(1)}_\mu=-\mu{\rm sin}(2q_2)\frac{2}{a}, \qquad Q^{(1)}_\rho=-\rho\nu_2q_1^2\frac{2}{a}, $$

and 
$$	\frac{d}{dt}\left(\frac{\partial L}{\partial \dot{q}_1}\right)-\frac{\partial L}{\partial q_1}+Q^{(1)}_{\alpha}+Q^{(1)}_{\mu}+Q^{(1)}_{\rho}=0,$$
becomes 
\begin{equation}
	\dot{q}_2=-\delta+(a^2\eta_1+\nu_1-\rho\nu_2)q_1^2-\mu{\rm sin}(2q_2)
\end{equation}

For $j=2$, a similar procedure yields

$$	\frac{d}{dt}\left(\frac{\partial L}{\partial \dot{q}_1}\right)=\frac{2}{a}\dot{q}_1, \qquad \frac{\partial L}{\partial q_1}=0,$$

$$Q^1_\alpha=\frac{4}{a}\alpha q_1, \qquad Q^1_\mu=-\frac{4}{a}\mu q_1{\rm cos}(2q_2), \qquad Q^1_\rho=\frac{8}{3a}\rho\kappa q_1^3, $$
and therefore 
\begin{equation}
	\dot{q}_1=-2\alpha q_1-\frac{4}{3}\rho\kappa q_1^3+2\mu q_1{\rm cos}(2q_2)
\end{equation}

\section{Fixed points}\label{app5}
Here I will show the derivation of the fixed point solutions.
The first fixed point condition, $\dot{q}_1=0$, leads to   
\begin{equation}\label{red_eq1b}
	q_1[2\alpha+\frac{4}{3}\rho\kappa q_1^2-2\mu{\rm cos}(2q_2)]=0,
\end{equation}
from where one obtains the trivial solution $q_1=0$ and the expression 
\begin{equation}\label{fix1}
	\mu{\rm cos}(2q_2)=\alpha+\frac{2}{3}\rho\kappa q_1^2
\end{equation}
The condition $\dot{q}_2=0$ yields 
\begin{equation}\label{fix2}
	\mu{\rm sin}(2q_2)=-\delta+(a^2\eta_1+\nu_1-\rho\nu_2)q_1^2.
\end{equation}

Combining Eqs.~(\ref{fix1}) and (\ref{fix2}), and using the condition 
${\rm cos}^2(2q_2)+{\rm sin}^2(2q_2)=1,$
one can obtain
$$(\alpha+C_1q_1^2)^2+(C_2q_1^2-\delta)^2-\mu^2=0,$$
and equivalently 
$$(C_1^2+C_2^2)q_1^4+2(\alpha C_1-\delta C_2)q_1^2+\alpha^2+\delta^2-\mu^2=0.$$
The solutions of this bi-quadratic equation are the non-trivial solutions $q_1^{(\pm)}$ defined through second expression in (\ref{sol_q1}).

Once $q_1$ is known, the contribution of the phase to the DS solutions can be obtained from Eq.~(\ref{fix1}) if one uses the equality ${\rm cos}(2 q_2)=1-2{\rm sin}(q_2)^2$ and ${\rm cos}(q_2)^2+{\rm sin}(q_2)^2=1$. With all this, the following expressions are obtained 
\begin{equation}
	{\rm sin}(q_2)=\sqrt{\frac{\mu-\alpha-2\rho\kappa q_1^2/3}{2\mu}},
\end{equation}
\begin{equation}
	{\rm cos}(q_2)=\sqrt{\frac{\mu+\alpha+2\rho\kappa q_1^2/3}{2\mu}}.
\end{equation}

\bibliographystyle{ieeetr}
\bibliography{ref}
\end{document}